\documentclass[aps,prl,preprint]{revtex4-2}
\usepackage{amsmath,amssymb,amscd,latexsym,amsthm}
\usepackage{cancel}
\usepackage{hyperref}

\begin{document}
\author{Peter Woit}
\affiliation{Department of Mathematics, Columbia University}
\email{woit@math.columbia.edu}
\title{Spacetime is Right-handed}
\begin{abstract}
We describe the relation between vectors and spinors in complex spacetime in an unconventional chirally asymmetric manner, using purely right-handed spinors, with Minkowski spacetime getting Wick rotated to a four-dimensional Euclidean spacetime with a distinguished direction.  In this right-handed spinor geometry self-dual two-forms can be used to get chiral formulations of the Yang-Mills and general relativity actions.  Euclidean spacetime left-handed spinors then transform under an internal $SU(2)$ symmetry, rather than the usual $SU(2)_L$ spacetime symmetry related by analytic continuation to  the Lorentz group $SL(2,\mathbf C)$.
\end{abstract}
\date{\today}
\maketitle
\newpage

\section{Wick rotation of four-dimensional vectors and spinors}
\label{sec:vector-spinor}

A simple way to understand Minkowski spacetime and its Lorentz symmetry is by identifying spacetime vectors $(x_0,x_1,x_2,x_3)$ with Hermitian two by two complex matrices
\begin{equation}
\label{eq:matrix}
\begin{pmatrix} x_0+x_3&x_1-ix_2\\x_1+ix_2& x_0-x_3\end{pmatrix}
\end{equation}
which have determinant minus the Minkowski length-squared: $-x_0^2+x_1^2+x_2^2+x_3^2$.
Elements $\Omega_L$ and $\Omega_R$ of the Lorentz group $SL(2,\mathbf C)$ of matrices of determinant one act by taking this matrix to
$$\Omega_L\begin{pmatrix} x_0+x_3&x_1-ix_2\\x_1+ix_2& x_0-x_3\end{pmatrix}\Omega_R^{-1}$$
preserving the Minkowski norm.  When $\Omega_R^{-1}=\Omega_L^\dagger$ the Hermiticity condition is preserved and one has an action on Minkowski spacetime.

The conventional approach to axiomatic relativistic quantum field theory (see e.g. \cite{streater-wightman}) allows the coordinates 
$x_j$ to be complex and describes expectation values of field operators (Wightman functions) as boundary values of functions holomorphic in the $x_j$.  Restricting such functions to real spatial coordinates and purely imaginary values of $x_0$ gives Schwinger functions of coordinates of a spacetime with Euclidean metric.   The analytic continuation between Wightman and Schwinger functions is what is often called \lq\lq Wick rotation".

If one allows all coordinates to be complex and drops the condition $\Omega_R^{-1}=\Omega_L^\dagger$, one gets complex spacetime with the complex Lorentz group $SL(2,\mathbf C)_L\times SL(2,\mathbf C)_R$ of pairs $(\Omega_L,\Omega_R)$ of $SL(2,\mathbf C)$ group elements acting on it as above.   A standard technique in axiomatic quantum field theory is to analytically continue in the coordinates of the complex Lorentz group between two of its real forms:
\begin{itemize}
\item
 the real Lorentz group $SL(2,\mathbf C)$ ($\Omega_R^{-1}=\Omega_L^\dagger)$
\item
the Euclidean rotation group ($\Omega_L,\Omega_R\in SU(2)$).  The action of such pairs preserves the real four-dimensional subspace with Euclidean metric ($x_0$ purely imaginary).
\end{itemize}
Taking Lorentz covariance of Wightman functions as an axiom, this analytic continuation implies $SU(2)_L\times SU(2)_R$ covariance of the Schwinger functions.

The group $SL(2,\mathbf C)$ has two inequivalent spinor representations:  the defining representation $(\frac{1}{2})$ on $\mathbf C^2$ and its conjugate $(\overline{\frac{1}{2}})$.  A coordinate invariant way to understand the description of complex spacetime in terms of matrices is to identify these matrices as linear maps from the dual of $(\frac{1}{2})_R$ to $(\frac{1}{2})_L$ (where $(\frac{1}{2})_{L,R}$ are the spinor representations of  $SL(2,\mathbf C)_{L,R}$).  Equivalently, complex spacetime is identified with the chirally symmetric holomorphic representation of the complex Lorentz group $SL(2,\mathbf C)_L\times SL(2,\mathbf C)_R$ on the tensor product 
\begin{equation}
\label{eq:usual}
\left(\frac{1}{2}\right)_L\otimes \left(\frac{1}{2}\right)_R
\end{equation}

In the van der Waerden notation, in Minkowski spacetime one writes coordinates for $SL(2,\mathbf C)$ $(\frac{1}{2})$ spinors as $s^A$ ($A=1,2$), and uses dotted  $s^{\dot A}$ for the conjugate $(\frac{1}{2})$ spinors. In complex spacetime conventionally one uses the same notation but now meaning: undotted indices for $(\frac{1}{2})_L\otimes (0)_R$, and dotted indices for $(0)_L\otimes (\frac{1}{2})_R$.  For more details see e.g. \cite{woodhouse}.

\section{A chiral alternative}

Instead of this chirally symmetric description, there is an alternate chirally asymmetric possibility that does not seem to have been previously considered.  This is to identify the Lorentz group not with a chirally symmetric set of adjoint pairs in $SL(2,\mathbf C)_L\times SL(2,\mathbf C)_R$, but with one of the factors, e.g. the right-handed $SL(2,\mathbf C)_R$.   The two by two matrices of equation \ref{eq:matrix} will then be linear maps from the dual of  $(\frac{1}{2})_R$ to $(\frac{1}{2})_R$, and 
complex spacetime will  be the non-holomorphic representation
$$(0)_L\otimes \left( \left(\frac{1}{2}\right)_R\otimes\left(\overline{\frac{1}{2}}\right)_R\right)$$

 If one just looks at the real Lorentz group, this is the same as the usual formalism.   But with this chirally asymmetric choice, there now is no $SL(2,\mathbf C)_L$ acting on complex spacetime.  This gives up the possibility of using the complex group $SL(2,\mathbf C)_L\times SL(2,\mathbf C)_R$ and a holomorphic representation to do analytic continuation, but one still has a complex spacetime in which one can analytically continue between real Minkowski and Euclidean subspaces.    

Twistor theory \cite{penrose} also involves a chirally asymmetric right-handed description of complex spacetime, with points in complex spacetime defined to be the $\mathbf C^2$ right-handed spinor space at that point, now a subspace of a $\mathbf C^4$ twistor space.  The left-handed spinors are the quotient of the twistor $\mathbf C^4$ by the right-handed spinor $\mathbf C^2$.   These two spinor spaces are the fibers of holomorphic bundles and the tangent bundle is conventionally the chirally symmetric holomorphic tensor product of the two kinds of spinors.   Complex spacetime in the twistor description again has Minkowski and Euclidean real subspaces, with vector and spinor fields related by analytic continuation.  The different possibility proposed here is that not just points but also the tangent bundle should be chirally asymmetric and purely right-handed.

\section{A distinguished imaginary time direction}

While $SL(2,\mathbf  C)$ acts on Minkowski space ($x_j$ real) in the usual way, something new happens on the Euclidean ($x_0$ pure imaginary) subspace.  $SU(2)_L$ acts trivially, the only non-trivial action is by $SU(2)_R\subset SL(2,\mathbf C)_R$.  The $(\frac{1}{2})_R$ and $(\overline{\frac{1}{2}})_R$ reps of $SL(2,\mathbf C)_R$ are equivalent as representations of $SU(2)_R$, and Euclidean spacetime is a real subspace of
$$\left(\frac{1}{2}\right)_R \otimes \left({\frac{1}{2}}\right)_R= (0)_R +(1)_R$$
Now there is a distinguished direction invariant under $SU(2)_R$ and a complementary three-dimensional subspace transforming as a vector under  $SU(2)_R$ (these are the real subspaces of the complex representations $(0)_R,(1)_R$).

Since the work of Schwinger \cite{schwinger} it has been apparent that a fundamental formulation of quantum field theory in Euclidean spacetime is possible and has many attractive features.  In particular, path integrals often are well-defined in Euclidean spacetime, not in Minkowski spacetime.   Euclidean quantum field theories have some very different properties than in the Minkowski case.  In particular, one can reconstruct the state space of the theory in Minkowski spacetime \cite{streater-wightman} in an $SL(2,\mathbf C)$ covariant way, with no distinguished time direction (one just needs to distinguish positive and negative energy).    Reconstruction of a physical state space from Euclidean spacetime Schwinger functions requires a choice of direction (one needs to restrict to positive imaginary time and use an Osterwalder-Schrader reflection, see \cite{osreconstruction}).    

In this right-handed complex spacetime version of Euclidean spacetime,  the fact that one has a distinguished direction in Euclidean spacetime appears naturally as part of the fundamental spinor geometry.  The fact that $SU(2)_L$ acts trivially, behaving like an internal  rather than spacetime symmetry opens up new possibilities for unification of interactions.  A proposed \lq\lq gravi-weak" unification, in which the Lorentz group is the right chiral half of an $SL(2,\mathbf C)_L\times SL(2,\mathbf C)_R$ symmetry, with the electroweak internal $SU(2)$ in the other half has been made in \cite{nesti-percacci} (see also \cite{alexander}, \cite{alexander-smolin}).  This proposal is set in Minkowski spacetime and does not involve Euclidean spacetime or twistor geometry. 

Given the necessity of the choice of a Euclidean spacetime vector pointing in the imaginary time direction in order to determine the physical state space, if one uses the conventional identification of Euclidean vectors and $\left(\frac{1}{2}\right)_L\otimes \left(\frac{1}{2}\right)_R$ spinors, one necessarily has a distinguished non-zero vector in the Euclidean real subspace of $\left(\frac{1}{2}\right)_L\otimes\left(\overline{\frac{1}{2}}\right)_R$.  This will provide an isomorphism between  $\left(\frac{1}{2}\right)_L$ and $\left(\overline{\frac{1}{2}}\right)_R$, allowing one to go back and forth between the chirally asymmetric description and the chirally symmetric one.

\section{Fundamental physics and right-handed Euclidean spacetime}

Remarkably, except for the Higgs field, all the fields of the Standard Model in Euclidean spacetime have chirally asymmetric descriptions and dynamics that only depend on the right-handed space-time degrees of freedom.  In this section we'll give a new such description of chiral spinor fields, and recall known such descriptions of gauge  and gravitational field degrees of freedom.

\subsection{Wick rotating Weyl spinor fields}

The Lagrangian density for a right-handed Weyl spinor degree of freedom in Minkowski spacetime is (in terms of momentum space fields) the $SL(2,\mathbf C)$ invariant
$$\mathcal L=\widetilde{\psi}^\dagger(p)(p_0-\boldsymbol \sigma\cdot\boldsymbol p)\widetilde{\psi}(p)$$
where $\widetilde{\psi}$ is in the $(\frac{1}{2})$ representation.  Solutions of the equation of motion have either
\begin{itemize}
\item
Positive energy $p_0$, so describe particles, and helicity $+\frac{1}{2}$.
\item
Negative energy $p_0$, so describe anti-particles, and helicity $-\frac{1}{2}$.
\end{itemize}
Here helicity is the eigenvalue of the operator
$$\frac{1}2{}\frac{\boldsymbol \sigma\cdot\boldsymbol p}{|\boldsymbol p|}$$

As noted for instance in \cite{ramond} (page 226),  with the conventional identification of  the Wick-rotated Euclidean spacetime vector $ip_0-\boldsymbol \sigma\cdot\boldsymbol p$ as an element of the tensor product $(\frac{1}{2})_L\otimes (\frac{1}{2})_R$, there is no way to get an invariant Lagrangian density just using right-handed Euclidean Weyl spinors.  One cannot construct a propagator for Euclidean spinor fields that transforms under $Spin(4)$ as expected without using both chiralities of spinor.  Even doing this, the conventional definition of Euclidean spinor fields involves an additional doubling of the number of degrees of freedom (see \cite{osterwalder-schrader}).

The formulation of Euclidean quantum field theory using a spacetime involving only right-handed spinor degrees of freedom avoids the Lagrangian density/propagator problem above without the necessity of introducing fields of the opposite (Euclidean) chirality.   The momentum space Dirac operator ($p_0-\boldsymbol \sigma\cdot\boldsymbol p$ in Minkowski spacetime, $p_0\rightarrow ip_0$ in Euclidean) involves the same identification of vectors with two by two complex matrices as discussed at the beginning of this article.  We have seen that one can interpret it as an element of the tensor product $\left(\frac{1}{2}\right)_R\otimes \left(\overline{\frac{1}{2}}\right)_R$, or use the more conventional interpretation as an element of  $\left(\frac{1}{2}\right)_L\otimes \left(\frac{1}{2}\right)_R$ together with a distinguished Euclidean spacetime direction vector which identifies the $\left(\frac{1}{2}\right)_L$ and $\left(\frac{1}{2}\right)_R$ Euclidean spinors.  

In \cite{hitchin} Hitchin defines the same sort of modified Euclidean Dirac operator, using the Clifford algebra basis element in the distinguished imaginary time direction.  He considers this operator coupled to a connection in a vector bundle, thus identifying the space of connections with a space of Dirac operators.  This infinite dimensional space has a hyperk\"ahler structure, with hyperk\"ahler quotient the space of (in his case anti)-self-dual connections.

Note that in Minkowski spacetime left-handed Weyl spinor fields can be constructed by conjugating right-handed Weyl spinor fields.   The chirally asymmetric right-handed description of spacetime takes both right-handed and left-handed Minkowski spinors to right-handed Euclidean spinors.

\subsection{Two-forms, spinors and Yang-Mills theory}
\label{sec:two-forms}

In four dimensions the Hodge dual takes two-forms to two-forms, satisfying $*^2=1$ in Euclidean signature.  Those with $*$ eigenvalue $+1$ are called self-dual, those with eigenvalue $-1$ are anti-self-dual.  Using the inner product, two-forms can be identified with infinitesimal $Spin(4)$ transformations, with self-dual two-forms corresponding to the Lie algebra of $SU(2)_R$ (so one could describe them as \lq\lq right-handed" instead of \lq\lq self-dual") and anti-self-dual ones corresponding to the Lie algebra of $SU(2)_L$.  In Minkowski signature one has $*^2=-1$, so eigenvalues are $\pm i$ and there are no real self-dual or anti-self-dual forms.  To get such forms, one needs to work with four complex dimensions and the complexified Lie algebras of $SL(2,\mathbf C)_R$ and $SL(2,\mathbf C)_L$.

Complexified self-dual two-forms are identified with the Lie algebra of $SL(2,\mathbf C)_R$, so are traceless maps from $(\frac{1}{2})_R$ to itself, so a three-complex dimensional space that can be identified with the symmetric part of the tensor product of two copies of  $(\frac{1}{2})_R$.  Like complex one-forms in the chirally asymmetric description of spacetime discussed above,  they  just involve the right-handed part of four-dimensional geometry. 

In the Euclidean version of Yang-Mills theory with gauge group $G$, the dynamical variables $A$ are connections valued in the Lie algebra of $G$, with dynamics given by the Yang-Mills action
$$\frac{1}{4g^2}|F_A|^2=\frac{1}{4g^2}\int Tr(F_A\wedge * F(A))$$
where $F_A$ is the curvature of $A$.  Decomposing $F$ into self-dual ($F^+_A$) and anti-self-dual ($F_A^-$) components, and using the fact that $\int Tr(F_A \wedge F_A)$ is a topological invariant, an equivalent form of the action is
$$ \frac{1}{2g^2}\int Tr(F_A^+\wedge * F_A^+)$$
This action only involves self-dual two-forms, and thus only right-handed spacetime spinor geometry.

\subsection{General relativity in terms of right-handed spinor variables}

It turns out that the dynamics of general relativity also can be formulated in a way that just involves right-handed spacetime.  That this simplifies the theory considerably was shown in \cite{ashtekar}, which described a new canonical formulation in terms of variable now known as Ashtekar variables.  These not only exploited the chiral description of four-dimensional spacetime, but also took as fundamental connection variables of the same sort as in Yang-Mills theory.  For a detailed treatment of chiral approaches to the formulation of general relativity, see \cite{krasnov}.  

One well-developed modern formalism for doing geometry and studying general relativity originates with Cartan.  In this formalism one works with objects defined equivariantly on the frame bundle of a manifold rather than the manifold itself.  This formalism allows a straightforward definition of spinor fields and is rather close to the formalism of gauge theory.   In gauge theory one starts with a principal $G$-bundle over a manifold $M$, with a connection and curvature described by $Lie\ G$-valued one and two-forms on $P$ satisfying an equivariance condition under the action of $G$ on $P$.  The gauge theory formalism is very general, allowing $G$ to be any Lie group, and it plays the role of an internal symmetry, unrelated to spacetime symmetry.

The Cartan formalism is a special case of the gauge theory formalism, taking $P$ as the bundle of orthonormal frames for a manifold with metric.  $G$ is then the orthogonal group acting on the frames.   One can do spinor geometry naturally in this formalism, working with a spin double-cover of $P$ and taking as group $G$ the spin double-cover of an orthogonal group.  In four dimensions and the Euclidean signature, this is Riemannian geometry with $G=SU(2)_L\times SU(2)_R$.  With Minkowski signature, $G$ is the Lorentz group $SL(2,\mathbf C)$.      The frame bundle of a manifold is a very specific  example of a principal bundle, with a group $G$ that is not independent of the spacetime geometry, but defined by it.   It comes with an additional structure one doesn't have in the gauge theory case, a canonical one-form $\theta$ on the frame bundle that takes values in $\mathbf R^4$.  This one-form is sometimes called the tetrad one-form.  At a point on the frame bundle, applied to a tangent vector, it gives the coordinates of the vector with respect to the frame.

When $G$ is the spin version of the frame bundle, the curvature two-form has remarkable properties, since it takes values in the Lie algebra of $G$, which is the same space as the space of two-forms.   So (at a point) one can think of it as an endomorphism  of a six-dimensional space with a decomposition into two three-dimensional spaces.  This has a block diagonal form, and the Einstein equations correspond to vanishing of the off-diagonal blocks (for details, see \cite{atiyah-hitchin-singer} or \cite{krasnov}).

Much of the effort to exploit Ashtekar variables or other chiral formulations of general relativity has been in Minkowski signature calculations, where one must complexify the tangent bundle and work with complex-valued connections and curvatures, then try to later impose reality conditions to get back to the real physical variables.   For a Lagrangian version of such an action for complex general relativity, see \cite{selfdual2forms}, where such a Lagrangian is given as
$$\int \Sigma^{\dot A\dot B}\wedge R_{\dot A\dot B}$$
Here we are using van der Waerden notation, with $\Sigma$ a self-dual two-form valued in the symmetric tensor product of two right-handed spinors that is constructed out of the one-form $\theta$.  $R_{\dot A\dot B}$ is the curvature two-form, taking values in the dual to the symmetric product of right-handed spinors.   Varying this action with respect to the connection gives the torsion-free (Levi-Civita) relation between the tetrad and the connection, while varying with respect to the tetrad gives the Einstein equations.

In Euclidean signature spacetime with Riemannian geometry one can work with real variables and does not need to complexify to get self-dual two-forms  There is a well-developed subject of Euclidean quantum gravity in which one attempts to start with a theory defined in Euclidean signature, and somehow later recover by analytic continuation the Minkowski signature theory.  This has well-known problems (see for instance \cite{gibbons}), but perhaps the chirally asymmetric right-handed version of Euclidean geometry discussed earlier might provide a different perspective.

\section{Discussion}

The Standard Model quantum field theory is not well-defined in Minkowski spacetime without some further information, with analytic continuation from Euclidean spacetime one way to accomplish this.  Defining the theory this way allows one to exploit the symmetries of the Euclidean spacetime, and  in \cite{woit} we described a speculative proposal for understanding the symmetries of the Standard Model in terms of the geometry of the Euclidean version of twistor space.  Such a twistor space description is inherently completely chirally asymmetric, with points in spacetime corresponding to spinors of one chirality.  The most unconventional part of the proposal is that the part of the Euclidean rotation group that acts on the other chirality could physically correspond not to a spacetime symmetry but to an internal symmetry.  The main goal here has been to understand the possible origin of such a counter-intuitive phenomenon.  Note that mixing between Euclidean rotational symmetry and an internal symmetry has been studied in other contexts, in particular in the twisting used to define topological quantum field theories (see \cite{witten-tqft}) and in formulating $N=4$ supersymmetry on the lattice (see section 8.2 of \cite{unsal}).
 
The conventional way of relating Euclidean and Minkowski spacetime spinors is by a chirally symmetric analytic continuation which takes Euclidean spacetime symmetries to Minkowski spacetime symmetries for both chiralities.  In this paper we have proposed a different relation, which uses just one chirality.   Spacetime (both Minkowski and Euclidean) can be said to be \lq\lq right-handed", and we have seen that this goes beyond the spin $\frac{1}{2}$ matter degrees of freedom, with Yang-Mills and gravitational dynamics also described using right-handed spinors.  The question of how exactly one can exploit the Euclidean spacetime $SU(2)_L$ symmetry that appears here to reproduce the usual electroweak theory requires further investigation.

\bibliographystyle{apsrev4-2}
\bibliography{righthanded}

\end{document}